\documentstyle[aps,prl,floats,graphicx,amssymb]{revtex}
\begin{document}
\draft
\twocolumn[\hsize\textwidth\columnwidth\hsize
\csname@twocolumnfalse\endcsname
\title{Theory of the $\pi$ state in $^3$He Josephson junctions}
\author{J.K. Viljas and E.V. Thuneberg} 
\address{Low Temperature Laboratory, Helsinki University of 
Technology, FIN-02015 HUT, Finland}
\date{\today}
\maketitle
\begin{abstract}
The flow of superfluid $^3$He-B through a $65\times 65$ array
of nanometer size apertures has been measured recently by Backhaus et
al. They find in the current--phase relation a new branch, so-called
$\pi$ state. We study two limiting cases which show that the
$\pi$ state arises from coupling of the phase degree of
freedom to the spin-orbit rotation. The $\pi$ state
exists in a single large aperture, but is difficult to
observe because of hysteresis. A better
correspondence with experiments is obtained by
assuming a thin wall, where the Josephson coupling
between the two sides arises from a dense array of pin-holes. 
\end{abstract} 
] 
\def\imagu{{\rm i}}
\def\re{\mathop{\rm Re}}
\def\im{\mathop{\rm Im}}

The flow of superfluid $^3$He-B through a single nanometer-size
aperture was studied by Avenel and Varoquaux some time ago \cite{AV}.
At temperatures near the superfluid transition temperature $T_{\rm c}$,
the current--phase relation is sinusoidal
\begin{equation}
 J(\phi)=J_{\rm c}\sin\phi
\label{e.jsin} \end{equation} 
as expected for a Josephson junction.
Also according to expectation, they find that the sine form
(\ref{e.jsin}) gradually becomes tilted when the temperature is
lowered. More recently, Backhaus et al.\ studied a $65\times 65$
array of small apertures \cite{packard,packard2}. They discovered a new
behavior where the current--phase relation acquires a positive slope at
phase differences $\phi\approx\pi$. This $\pi$ state develops when the
temperature is lowered to approximately $0.6T_{\rm c}$. 

A few theoretical explanations for the $\pi$ state have been proposed
\cite{hatakenaka,AMV}. In this letter we present a theory, that
is based on the many-component form of the order parameter in $^3$He.
It differs from the previous suggestions because it contains
no unjustified assumptions and an order-of-magnitude agreement with
experiments is obtained without any adjustable parameters.

Unusual current-phase relations occur also in  other
systems. A $\pi$ junction, where $J_{\rm c}$ in (\ref{e.jsin}) is
negative, can be induced by adding magnetic impurities to a
tunneling barrier between two s-wave superconductors \cite{BKS}. Similar
$\pi$ shifts appear in nonmagnetic junctions between d-wave
superconductors. In addition,  current-phase relations with additional
zeros ($J(\phi)=0$ for $\phi\not= 0$ or $\pi$) can appear for
special orientations of the anisotropic crystals
\cite{yip}. The $\pi$ state in $^3$He differs from these
in several respects, most fundamentally because it
arises from interplay of two soft modes of the order
parameter, the phase $\phi$ and the spin-orbit
rotation. 

We present calculations in two limiting cases. In the
case of a tunneling barrier, the existence of the $\pi$
state can be demonstrated by analytic calculations. 
The parameters of the tunneling model are estimated
using the quasiclassical theory.  In the case of a single
aperture, the $\pi$ state is obtained by numerical
simulations using the Ginzburg-Landau theory of
$^3$He.

{\it Tunneling junction.}---The simplest case to demonstrate the $\pi$
state is to consider a planar wall through which the $^3$He
atoms can tunnel. The energy arising from tunneling 
between the left ($L$) and right ($R$)
sides can be written as \cite{AGR}
\begin{equation}
 F_{\rm J}=- \re \sum_\mu\left[a A_{\mu z}^{L*}A_{\mu z}^R
 +b(A_{\mu x}^{L*}A_{\mu x}^R+A_{\mu y}^{L*}A_{\mu y}^R)\right].
\label{e.tunnelingF0}\end{equation}
Here $A_{\mu j}$ is the $3\times 3$ matrix order parameter where the
first index $\mu$ refers to the orientation of the Cooper pair in
spin space, and the latter index $j$ in orbital space. The $z$ axis is
taken perpendicular to the tunneling wall. Equation
(\ref{e.tunnelingF0}) is a simple generalization of
$F_{\rm J}=-a\re (A^{L*}A^R)$, which describes the Josephson coupling
of two s-wave superconductors with order parameters $A^{L}$ and $A^R$
\cite{Tinkham}. The mass current through the wall is given by
$J=(2m_3/\hbar)\partial F_{\rm J}/\partial\phi$, where
$\phi=\phi^L-\phi^R$ is the phase difference and $m_3$ the mass
of a $^3$He atom.

In the B phase of $^3$He, the order parameter has the form 
$A_{\mu j}=\Delta\exp(\imagu\phi)R_{\mu j}$. Here $\Delta$ is the
amplitude, $\exp(\imagu\phi)$ a phase factor, and $R_{\mu j}$ a
$3\times 3$ rotation matrix: $\sum_\mu R_{\mu j}R_{\mu k}=\delta_{jk}$.
The rotation matrices can be parametrized by an axis $\hat{\bf n}$ and
an angle $\theta$. Substituting into (\ref{e.tunnelingF0}) gives
($\alpha$, $\beta>0$)
\begin{equation}
 F_{\rm J}=-\sum_\mu \left[\alpha R_{\mu z}^LR_{\mu z}^R
 +\beta(R_{\mu x}^LR_{\mu x}^R+R_{\mu y}^LR_{\mu y}^R)\right]\cos\phi.
\label{e.tunnelingF}\end{equation}
In deriving this expression from (\ref{e.tunnelingF0})
one must pay attention to the fact that the  order
parameter of the p-wave superfluid is strongly
suppressed near a wall. As a consequence the
parameters $\alpha$ and $\beta$ in 
(\ref{e.tunnelingF}) are not simply related to the
coefficients $a$ and $b$ in (\ref{e.tunnelingF0}), but
otherwise the dependence of $F_{\rm J}$ on the soft
variables $\phi$ and $R_{\mu j}$ remains the same as
obtained by the simple substitution above.

\begin{figure}[bt] 
\begin{center}\leavevmode
\includegraphics[width=0.8\linewidth]{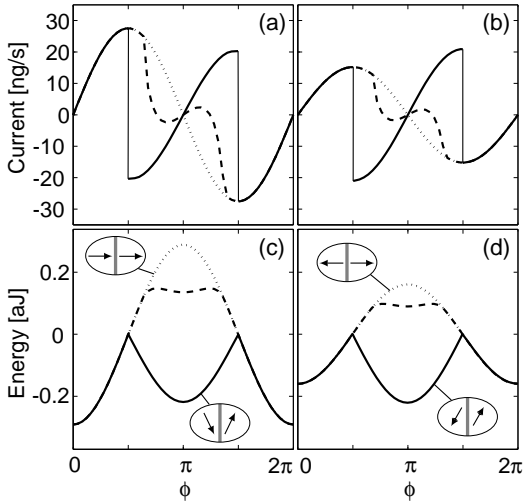}
\bigskip \caption{   The current--phase relationships and
energies for the tunneling model. The left and right panels correspond 
to parallel and antiparallel $\hat{\bf n}$'s far away from the junction,
respectively. The directions near the junction are depicted by arrows.
The curves correspond to different values of the gradient-energy
parameter $\gamma$: ideal $\pi$ state  ($\gamma=0$,
solid line), no $\pi$ state  ($\gamma=\infty$, dotted
line), and an intermediate case ($\gamma=0.245$ aJ, dashed
line). The parameters $\alpha=0.2207$ aJ and
$\beta=0.0347$ aJ are chosen to imitate the 
experiment \protect\cite{packard2} at $T=0.55T_{\rm c}$.    
 }\label{f.gamma}\end{center}\end{figure}

Let us consider the case that the rotation matrices on the left and
right sides are the same. This gives rise to the
``zero-state'' with the critical current $J_{\rm
c}=(2m_3/\hbar)(\alpha+2\beta)>0$. This state has
lowest energy when $\vert\phi\vert<\pi/2$ because it
corresponds to the maximum of the expression in
square brackets in (\ref{e.tunnelingF}). The situation
changes when $\phi$ exceeds $\pi/2$. There one has to
look for a {\it minimum} of the expression in the square
brackets. This corresponds to the $\pi$ state, which is
illustrated by the solid line in Fig.\ \ref{f.gamma}a.
The critical current $J_{\rm c}$ in (\ref{e.jsin}) is
negative:  $J_{\rm c}=-(2m_3/\hbar)\alpha$ if
$\alpha>\beta$ and  $J_{\rm
c}=-(2m_3/\hbar)(2\beta-\alpha)$ otherwise.

In order to make the tunneling model realistic, we have to consider
three additional contributions to the energy. Firstly, there is the
magnetic dipole-dipole energy $F_{\rm d}=8g_{\rm d}\Delta^2({1\over
4}+\cos\theta)^2$
\cite{Leggett}. In the bulk it fixes the rotation angle
$\theta$ equal to $\theta_0=\arccos(-{1\over
4})=104^\circ$. This remains valid also near the junction because both
the Josephson energy (\ref{e.tunnelingF}) and the dipole-dipole energy
can reach their minima simultaneously: the products of
two rotation matrices appearing in the former are not limited by the
fact that both matrices have a fixed
rotation angle $\theta_0$. Secondly, there is a surface
energy that arises from coupling of the dipole-dipole energy to the
suppression of the order parameter near walls \cite{BSOB}. It has the
form
\begin{equation} F_{\rm s}=
b_4(\hat{\bf n}\cdot\hat{\bf s})^4-b_2(\hat{\bf n}\cdot\hat{\bf s})^2,
\label{e.fsurf}\end{equation}
where $\hat{\bf s}$ is the surface normal. The lowest surface
energy is achieved when the rotation axis $\hat{\bf n}$
is perpendicular to the wall,
$\hat{\bf n}=\pm\hat{\bf s}$, because
$b_2>2b_4>0$. Thirdly, there is a gradient energy
associated with spatial bending of the rotation axis $\hat{\bf n}$.
It arises because in practice all tunnel junctions are
of finite size, and other walls in the container favor
a different orientation of $\hat{\bf n}$ than may be the
minimum of the Josephson energy. We model the
gradient energy by the simple quadratic forms \begin{equation}
F_{\rm g}^L=\gamma(\eta^L-\eta_\infty^L)^2,\ 
F_{\rm g}^R=\gamma(\eta^R-\eta_\infty^R)^2,
\label{e.grad}\end{equation} where $\eta$ is the polar
angle of $\hat{\bf n}$, {\it i.e.}, $\cos\eta=\hat n_z$.
$\eta^L$ and $\eta^R$ denote the polar angles on both
sides just at the junction, and we assume that the
values $\eta_\infty^L$ and $\eta_\infty^R$ further
away are either $0$ or $\pi$. In the experimental case
\cite{packard} the surface energy
(\ref{e.fsurf}) is important in fixing $\eta_\infty^L$
and $\eta_\infty^R$, but otherwise its contribution is
so small that we can neglect it in the following.   

The current--phase relations for the tunneling model,
(\ref{e.tunnelingF}) and (\ref{e.grad}), are
plotted in Fig.\
\ref{f.gamma}. It can be seen that a large value of the gradient
energy parameter $\gamma$ suppresses the $\pi$ state. Furthermore, we
find two cases where the rotation axes $\hat{\bf n}$ far from the
junction are either parallel or antiparallel. The latter has smaller
critical current [$J_{\rm
c}=(2m_3/\hbar)(\alpha-{7\over 4}\beta)$] but a relatively a more
pronounced
$\pi$ state. This ``bi-stability'' was theoretically discussed in 
Ref.\ \cite{T88} and has
recently been observed experimentally \cite{packard2}. 

\begin{figure}[bt]
\begin{center}\leavevmode
\includegraphics[width=0.9\linewidth]{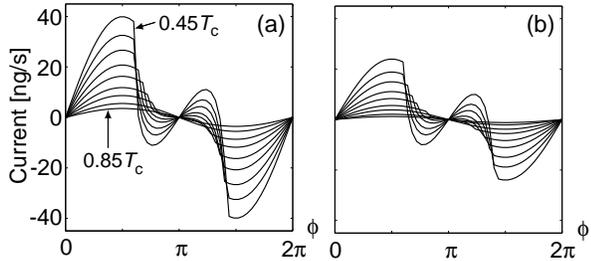}
\bigskip
\caption{The current--phase relationships for parallel (a) and
antiparallel (b) $\hat{\bf n}$'s far away from the junction.
The different curves correspond to temperatures from $0.45T_{\rm c}$ to
$0.85T_{\rm c}$ with intervals of
$0.05T_{\rm c}$. The parameters $\alpha$ and $\beta$ are calculated
with the pin-hole model and $\gamma$ estimated as explained in the text.
However, in order to get better correspondence with experiments, we
have multiplied the estimated values by factors $7.5$,
$1.3$, and $0.15$, respectively. 
 }\label{f.ctemp}\end{center}\end{figure}

{\it Evaluation of tunneling parameters.}---The experiment
\cite{packard} has a square array of apertures of diameter
$D=100$ nm with spacing $S=3\ \mu$m in a wall of thickness $W=50$ nm.
In order to make the tunneling model to imitate the experiment, we
estimate
$\alpha$ and $\beta$ by letting all the three lengths to approach zero
but keeping their ratios unchanged \cite{note2}. The calculation for
such ``pin-holes'' \cite{kurkijarvi} is relatively simple once the
self-consistent solution for the order parameter near a wall is known
\cite{ZKT}.  We assume diffuse scattering of quasiparticles at 
surfaces. The tunneling form (\ref{e.tunnelingF}) is
reproduced in these calculations at temperatures
$T\gtrsim 0.5T_{\rm c}$, and values of $\alpha$ and
$\beta$ can be extracted. The parameter $\gamma$ can be estimated using
the bending energy of the B phase \cite{Cross} and
assuming a simple form $\eta({\bf
r})-\eta_\infty\propto r^{-1}$, where $r$ is the radius from the
center of the aperture array, and this expression is cut off at the
radius of the array. In agreement with experiments, we find that the
$\pi$ state appears at low temperatures because $\alpha$ and
$\beta\propto(1-T/T_{\rm c})^2$ have stronger
temperature dependence than $\gamma\propto
1-T/T_{\rm c}$, see Fig.\ \ref{f.ctemp}. Moreover, the parameters
$\alpha$, $\beta$, and $\gamma$ agree within one
order of magnitude to those that give an approximate best fit to the
experiments (see caption of Fig.\ \ref{f.ctemp}). This fit reproduces
also the absolute magnitude of the critical current, and the 
same values of the parameters are used for cases of both parallel and
antiparallel
$\hat{\bf n}$'s. 

The tunneling model can be improved trivially by extending the
pin-hole calculation to the whole temperature range $0<T<T_{\rm c}$.
A more ambitious project for the future would be the self-consistent
calculation for aperture sizes on the order of the coherence length
$\xi_0$. In both cases the resulting Josephson energy 
$F_{\rm J}(\phi, R_{\mu j}^L, R_{\mu j}^R)$  will no more be of the
simple form (\ref{e.tunnelingF}). 

{\it Single aperture.}---The limit opposite to the
tunneling barrier is a single large aperture. There the
major task is to calculate the order parameter
self-consistently. We have done this using the
Ginzburg-Landau (GL) theory of $^3$He. The differential
equations were solved numerically on a grid in and around the aperture.
Our calculations are more general than the previous ones
\cite{T88,ullah} because we use a full a three-dimensional grid.
Vanishing $A_{\mu j}$ was assumed at surfaces.

\begin{figure}[bt]
\begin{center}\leavevmode
\includegraphics[width=0.8\linewidth]{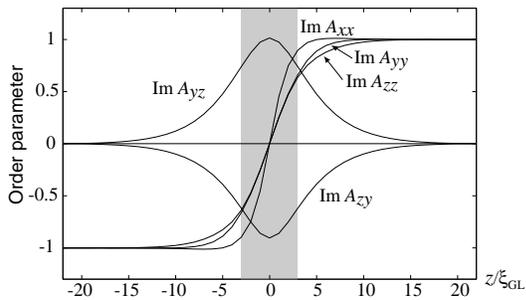}
\bigskip
\caption{  
The order parameter in the $\pi$
state along the axis $z$ of the aperture.  The figure corresponds to
the phase difference $\phi=\pi$, i.e., the order parameters far left
and far right differ by factor -1. The broken symmetry allows nonzero
$A_{yz}$ and
$A_{yz}$ which have a long tail in the bulk. These components vanish
in the zero branch.
The wall is shown as shaded, $W= 6\xi_{\rm GL}$ and $D= 10
\xi_{\rm GL}$. 
 }\label{f.piop}\end{center}\end{figure}

The order parameter of the $\pi$ state is shown in Fig.\ \ref{f.piop}.
It is plotted along the axis of a circularly symmetric
aperture. For simplicity, we have normalized the order 
parameter to unit matrix in the bulk,  $A_{\mu
j}(z=\pm\infty)=\exp(\pm{\rm i}\phi/2)\delta_{\mu j}$
(assuming the case of parallel $\hat{\bf n}$'s). This is
possible  because for aperture sizes on the order of
the GL coherence length $\xi_{\rm GL}$ the
dipole-dipole energy can be
neglected. The characteristic property of the $\pi$
state is the components $A_{yz}$ and $A_{zy}$. These
are the dominant components in the orifice, and they
decay slowly towards the bulk. They imply
broken symmetry: the symmetry group of the $\pi$
branch is $m'm2'$ compared to ${\infty\over
m'}{2'\over m}$ of the zero branch. (Here prime denotes
time-inversion.) This sets rather strong requirement for
the calculation because the circular symmetry of the
aperture cannot be used to simplify the
computation.      

\begin{figure}[bt]
\begin{center}\leavevmode
\includegraphics[width=0.8\linewidth]{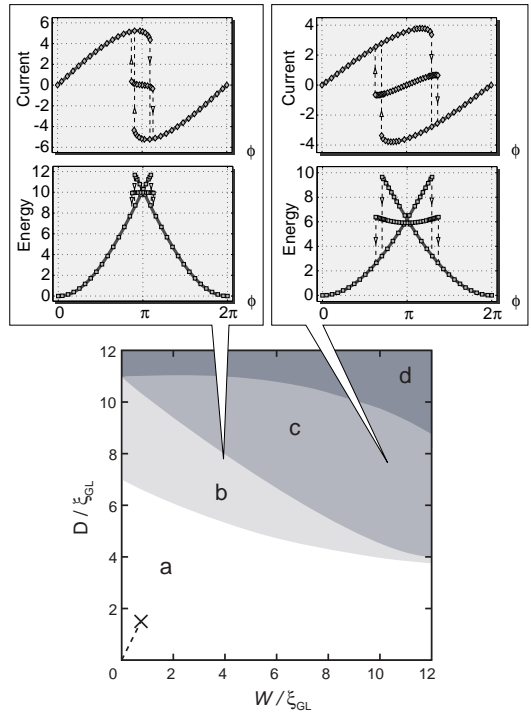}
\bigskip
\caption{  
Theoretical phase diagram for the the $\pi$ state in a single 
aperture. The current and the energy as a function of phase difference 
$\phi$ are shown in two cases.
$D$ is the diameter of the aperture and $W$ the wall thickness. The
$\pi$ branch is found in regions (b), (c), and (d), where it is
locally stable at a fixed phase difference $\phi\sim\pi$. In
regions (c) and (d) it is also a local minimum of
energy with respect to $\phi$ at $\phi=\pi$. In regions
(b) and (c) the $\pi$ state is the absolute minimum
energy state at $\phi=\pi$. The parameters of {\it one}
aperture of the experimental aperture array
\protect\cite{packard} are shown by dashed line, and
the observation of the $\pi$ state is marked by a cross.
The temperature dependent GL coherence length is
defined by $\xi_{\rm GL}=\hbar v_{\rm
F}/\sqrt{10}\Delta(T)$, where $v_{\rm F}$ is the Fermi
velocity and $\Delta(T)$ the BCS energy gap. 
}\label{f.josgl}\end{center}\end{figure}

The current--phase relations are summarized in Fig.\
\ref{f.josgl}.
We see that the occurrence of the $\pi$ branch depends
sensitively on the diameter of the aperture, whereas
the wall thickness is less important. For small
apertures no $\pi$ branch is found. When $D$ exceeds approximately 
$5 \xi_{\rm GL}$, the $\pi$ branch appears.  In the
region (b) the current--phase relation has negative
slope. Such a state can be stabilized if the left and right
sides are connected, like in a torus geometry.
Increasing the diameter, the current--phase relation
gets a positive slope in region (c). This state is stable
also in a piston-driven flow channel. The $\pi$ state is
also the absolute energy minimum at $\phi=\pi$ in
region (c). In region (d) the $\pi$ state continues to exist but it
has higher energy than the zero branch. The calculation assumes the
idealized case of flow between two infinite bulk fluids. Any additional
hydrodynamic inductance
 shifts upwards the border between regions (c) and (d) 
as it increases the energy of current  carrying
states.

Although the $\pi$ branch constitutes the absolute energy minimum, it
may be difficult to find it experimentally in a single aperture. The
reason is that whenever it is locally stable, there always exists a
locally stable zero state at the same $\phi$. Because the order
parameters of the $\pi$ and zero states differ considerably, it may be
that the phase slips only take place between two branches of the zero
state without ever finding the way to the lower energy $\pi$ state.
This is what we find in the numerical calculations, where the $\pi$
state was found only if the initial $A_{\mu j}$ was chosen close enough
to the converged solution. We remind, though, that our calculations are
not meant to simulate the correct dynamics of the phase slip. 

The dimensions of one  aperture in the array at Berkeley 
  are marked on figure \ref{f.josgl}. This is clearly in the region
where no $\pi$ state is found. Although the Ginzburg-Landau calculation
is accurate only at temperatures near $T_{\rm c}$, it is unlikely that
the $\pi$ state could be stabilized in a more accurate calculation at
lower temperatures.  This statement is based on
experience gained in 
previous low-temperature calculations \cite{TKS}. Thus
we conclude that the appearance of the $\pi$ state in
the Berkeley experiment essentially depends on the
presence of many apertures.  

We have also done two-dimensional calculations that simulate the flow
through a long narrow slit. The $\pi$ state is found and its properties
are qualitatively similar to those in a circular aperture
\cite{note}. In particular, the transitions from the zero branch to the
$\pi$ branch seem to be absent. This is consistent with the fact that
no $\pi$ branch was found in the experiments by Avenel and Varoquaux
\cite{AV}. 

The $\pi$ state can be interpreted so that a {\it half-quantum} vortex
has crossed the orifice. There are no free half quantum vortices in
superfluid $^3$He-B, but the double-core vortex \cite{T87} can be
interpreted as a bound pair of two half-quantum vortices \cite{hqv}.
Indeed, the order parameter in Fig.\ \ref{f.piop} is very similar to
that in the double-core vortex on the axis going between the two
cores \cite{T87}. 

{\it Conclusion.}---The $\pi$ state was found to occur in both the
limits investigated above. Its mechanism is the
same in both cases: a lower coupling energy is
achieved by producing a spin-orbit rotation that heals
slowly in the bulk liquid.  The Berkeley experiment
\cite{packard} is somewhat an intermediate case to the
two limits studied, which makes it evident that the $\pi$ state
also there arises  from the same mechanism. 

The theory above provides several  predictions that
can be tested experimentally. For example, the  
$\pi$ state depends on the linear dimension $L$ of the aperture array
because $\alpha/\gamma$, $\beta/\gamma\propto L$. An external
magnetic field fixes the surface orientation of $\hat{\bf n}$, and
thus can be used to suppress the $\pi$ state.


\begin{thebibliography}{9}
\bibitem{note2} Note that this procedure limits the grazing-angle
quasiparticles more than the cosine law usually employed in pin-hole
calculations \protect\cite{kurkijarvi}.
\end{thebibliography}
\end{document}